# Cascading crystallographic transitions α → ω → β → β' → β" and melting curve of elemental zirconium

Joseph Gal*
Ilse Katz Institute for Nanoscale Science and Technology ,
Ben-Gurion University of the Negev, Beer Sheva ,84105 Israel

## Abstract

Precise fitting of the experimental data analyzed separately for each identified crystallographic phase (α, ω, β, β' and β") yield different bulk moduli $B_o$ and $B_o'$ and different zero pressure volumes ($V_o$) than those claimed in the literature. Special attention is given to the bcc phases indicating cascading transitions β → β' → β" associated with volume collapse. The present analysis reveals the existence of a bcc-β' phase which is reported here for the first time. It is shown that the first order volume collapse at ~58GPa (β → β') is followed by a moderate transition to the bcc-β" phase. The β' phase is stable up to 110GPa. Above 110GPa the bcc-β" is dominant and stable up to ~220GPa. The derived bcc-β" bulk moduli are confirmed by the Lindemann-Gilvarry criterion as $B_o$ and $B_o'$ simultaneously fit both the P-V EOS and the P-T melting data points (combined approach). The calculated melting curve of elemental Zr, taking into account the thermal pressure $Po_{th}$ shift and the elevated melting $Tm_o'$ at $Po_{th}$, yield very good fit of the experimental melting data permitting a safe extrapolation to high pressures and temperatures. In addition, the combined approach lead to direct determination of the Grüneisen parameter $\gamma_o$, needed for applying the approximated Lindemann-Gilvarry melting formula.
It is shown the DFT(QMD) simulations and the non-hydrostatic thermodynamic formalism for solid mediums mismatch the experimental melting data.

*jgal@bgu.ac.il

*Introduction*- The physical-mechanical and melting properties of zirconium metal are of particular interest for the nuclear industry because of Zr's low neutron absorption cross section and relatively high melting temperature (2128K). In addition, the bcc-to-bcc (β → β') isostructural phase transition associated with volume collapse has attracted the scientific community during the last two decades. Pressure induced first-order isostructural transitions associated with volume collapse had been observed only in Ce, where an isostructural fcc→fcc phase transition with a substantial volume decrease up to a critical point is a well known effect [2,3]. The exact origin of the isostructural transition in Ce is still under debate, the general consensus, however, is that this transition is driven by a change in the degree of the localization and correlation of the one 4f electron. As early as in 1991 an isostructural bcc→bcc phase transition at ~58 GPa in elemental Zr was suggested by Akahama et al. [4], speculating that this transition is triggered by s-d electronic transition.

Recently, Stavrou et al. [1] have performed a precise XRD structural study of Zr under pressure up to 210GPa at ambient temperature. Reconfirmation of the existence of an isostructural bcc-to-bcc phase transition was reported, claiming a first-order transformation. In addition, the bulk moduli of the β and β' phases were derived utilizing the Birch-Murnaghan (BM) equation of state (EOS) [5]. Unfortunately, the fitting curves are not displayed in Fig.1 of Ref.1 and their results are only summarized there in Table I. In the present contribution I have performed accurate fittings of this reported data using Vinet [6] and BM EOS. The fitting results revealed the existence of another bcc phase with different elastic properties in addition to the reported β and β' phases.

The melting temperatures up to ~80 GPa of Zr metal confined in a laser heated diamond anvil cell (DAC) were recently reported by Parisiades et al. [7]. Poor fittings of the experimental data were obtained by introducing the Lindemann-Gilvarry (LG) criterion and Simon-Glatzel procedure which are depicted in the proposed phase diagram (there in Fig.5), the reason of which will be explained below.

In two recent publications [8,9] I have claimed that isochoric conditions exist in the DAC and the LG criterion [10] is applicable for predicting the melting curves of metals. By introducing a constraint demanding that the fitting of the experimental EOS (P-V space) data will simultaneously fit the experimental melting (P-T space) (combined approach), consistent bulk moduli and melting curves are obtained.

The LG criterion is not a theoretical model based on first principles but a phenomenological approach to the behavior of solids. Nevertheless, it is a solid fact that for several metals the combined approach provide good fits of the melting experimental data at high pressures and temperatures [8,9]. The LG criterion needs the Grüneisen parameter $\gamma$. The procedure using the LG criterion together with the Grüneisen parameter $\gamma$, according to the Slater model [11] often does not fit the experimental melting data [12]. The LG formulation uses the bulk modulus B and its pressure derivative B' as fit parameters deduced directly from the EOS. However, the fitting results of the EOS are not unique reliant on the chosen EOS as well as on the pressure transmitting medium (PTM). The EOS [5,6] need two free parameters; the bulk moduli B and B' which are deduced from the P-V room temperature isotherm and are assigned $B_o$ and $B_o$'. Therefore, the reported values of $B_o$ and $B_o$' for aluminum [13,14,15], copper [15,16], and uranium [18, 19,20,21,22] range up to ~ 50%. In β-Zr the bulk moduli spread from 79 to 255GPa according to analysis reported by Greef [17], Stavrou [1] and Akahma [4]. Thus, the question remains, which of these bulk moduli should be addressed?

In the present paper the experimental data of β-Zirconium β" phase is reanalyzed using the combined approach, allowing to obtain reasonable bulk moduli and enabling the extrapolation the melting curve to high pressures and temperatures.

**_Lindemann-Gilvarry approximation_**- According to Lindenmann's criterion
The melting temperature $T_m$ is related to the Debye temperature $\Theta_D$ as follows:

$$T_m = C \, V^{2/3} \, \Theta_D^2 \qquad (1)$$

Where V is the volume and C is a constant to be derived for each specific metal. In the Debye model the Grüneisen parameter $\gamma$ is defined by $\gamma = \partial \ln \Theta_D / \partial \ln V$. As shown by Anderson and Isaak [4] combining (1) and (2), inserting $V_o/V = \rho/\rho_o$, and integrating one gets the form of the LG criterion for the melting temperature $T_m$ :

$$T_m(\rho) = T_{mo} \exp \left\{ \int_{\rho o}^{\rho} [2\gamma - 2/3] \, d\rho/\rho \right\} \qquad (2)$$

Where $\rho_o$ is a reference density, $\rho$ is the density at the melt and $T_{mo}$ is the melting temperature at the reference density. Integrating (2) assuming that $\gamma = \gamma_o \, (\rho_o/\rho)^q$ and q=1 one gets:

$$T_m(V) = T_{mo} \, (\rho_o/\rho)^{2/3} \, \exp[2\gamma_o (1 - \rho_o/\rho)] \qquad (3)$$

where $\gamma_o$ is defined as the Grüneisen parameter at ambient conditions [10].

Equation (3) states that if $\rho(P)$, $T_{mo}$ and $\gamma_o$ are known the melting curve $T_m(P)$ can simply be determined assuming that the relation between P and $\rho$ is known.

It is well accepted that the pressure in the P-V-T space is given by:

$$P(V,T) = P_C + \gamma_{lattice} \, C_{v \, lattice} \, \rho \, [T - T_o + E_o/C_{v \, lattice}] + \tfrac{1}{4} \rho_o \gamma_e \beta_o (\rho/\rho_o)^{1/2} T^2 \qquad (4)$$

Here Pc is the cold pressure, $C_v$ is the lattice specific heat above $T_o$, $T_o$ is the ambient temperature. $C_{v,lattice}$ is taken as constant (usually at room temperature, following the approximation of Altshuler et al. [13] ) , $E_o$ is the lattice thermal energy at $T_o$ and $\gamma_{lattic}$ is the lattice Grüneisen parameter. $\gamma_e$ is electronic Grüneisen parameter and $\beta_o$ is the electronic specific heat coefficient (Altshuler [13] and Kormer [24]). In most experiments, the material is compressed at room temperature and then heated to the melting point. The results obtained are known as the cold melting curve forming Pc. In the second stage the actual pressure is obtained by adding the calculated thermal pressure demanding that the shock wave data should serve as anchor to the fitting procedure. This two stage procedure have been confirmed for Al,Cu and U melting curves in reference [8].

The relation between $P_C$ and the density $\rho(P)$ for the room temperature isotherms are given by Vinet [5] or third order Birch-Murnaghan [6] equations of state. $\rho$ is density and $B= -V (\partial P/\partial V)$ is the definition of the bulk modulus and B' is the pressure derivative of the bulk modulus ( $B'=\partial B/\partial P$). B and B' are fit parameters of the room temperature isotherm assigned as $B_o$ and $B_o$'. As stated in the introduction, the best fit solutions are not unique and occasionally depends on the chosen EOS. This is the reason why diverse results are obtained by different authors.

Thus, it make sense to introduce a different procedure in order to improve the fittings of the data in the P-V and the P-T planes.

The following four step procedure to determine the correct melting curve (the combined approach) was proposed in [8]:

1. Utilize Lindemann-Gilvarry criterion (eq.3) with $\gamma_{eff}$ as a free parameter [8] and optimize $B_o$ and $B_o$' by choosing the appropriate EOS which best fit simultaneously, the experimental P-V data (isotherm 300K) and the experimental melting P-T data. In this way $P_c$ is obtained, forming the cold melting curve. In LG eq.3 $Tm_o$ and $V_o$ are the melting temperature and volume at ambient pressure.

2. Adding the calculated thermal pressure $P_{th}$ to $P_c$ obtaining the LG melting curve accounting for the actual pressure (isochoric condition) sensed by the investigated sample. Demanding that the thermally corrected melting curve will include the shock wave melting data as anchor. The Grüneisen parameter $\gamma_o$ is derived accordingly.

   In the present Zr case the calculated thermal corrected melting curve should be applied [7].

3. Extrapolating the derived thermally corrected melting curve to high pressures and temperatures.

In DAC experiments, the material is compressed at room temperature and then heated to the melting point. In this case the claimed pressure is not the actual pressure and the $\gamma_o$ should be replaced by $\gamma_{eff}$ in eq.3 when fitting the as measured experimental data, thus forming the cold melting curve. By applying this procedure safe extrapolation of the melting curves to high pressures and temperatures done for Al, Cu, or U metals [8].

In a well constructed DAC with the proper transmitting medium isochoric condition exist, thus, upon heating the sample confined in the cell, thermal pressure develops associated with the increase of the melting temperature. The calculated thermal pressure ($Po_{th}$) and the melting point $Tm_o$' at ambient pressure are derived by calculating $Po_{th}$ according to eq.4 and adjusting $\gamma_o$ to match the shock wave data forming the actual melting curve. To clarify, $Po_{th}$ is the pressure shift from ambient pressure and $Tm_o$' is the melting temperature at $Po_{th}$. For Zr the calculated shift is 8GPa and $Tm_o$'=2300K. The textbook melting point 2128K at zero pressure as derived in the open space simply does not exist in the heated DAC as depicted in Fig 2.

**Results-** The precise DAC measurements of the EOS elemental zirconium at ambient temperature (isotherm 300K) as reported by Stavrou et al. in PRB (2019) [1] is the first measurement performed with no pressure transmission media. In other words, Zr metal fine powder is the PTM preventing uniaxial stress leading to a reliable EOS.

The analysis of this data using VIN or BM EOS is depicted in Fig.1 (solid lines). The experimental data of the hcp-α phase best fitted with the VIN EOS suggest bulk moduli $B_o$= 160GPa and $B_o$'= 1 and $V_o$=23 Å$^3$/at. (red solid line). The best fit of the hex-ω phase is obtained with the BM EOS suggesting $B_o$= 190GPa and $B_o$'= 3.3 and $V_o$=25 Å$^3$/at. (green solid line). Increasing the pressure to above 34GPa the hexagonal crystallographic phase transforms to the bcc-β phase. Further increasing the pressure reveal cascade of transitions to bcc phases, namely to β' and β". The β phase is fitted with VIN EOS with bulk moduli parameters $B_o$=203(20) GPa and $B_o$'=3.3(2) in accord with Akahama [4]. Extrapolating to zero pressure reveals $V_o$ = 20(1) (Å$^3$/at.) (solid brown line). The second phase assigned β' is fitted with VIN EOS and $B_o$=230(20) GPa and $B_o$'=2.0(2) parameters, extrapolation to zero pressure lead to $V_o$ = 18.3(2) (Å$^3$/at.) as the best fit. The third phase β" is fitted simultaneously (combined approach steps 1and 2) with VIN EOS the bulk moduli parameters $B_o$=265(10) GPa and $B_o$'=3.3(1) where $V_o$ = 16.8(2) (Å$^3$/at.). However, the BM EOS reveal $B_o$=255(5) and $B_o$'=3.3(1) and $V_o$ = 21.3(2) (Å$^3$/at.) in agreement with $B_o$ claimed in [1].

The cascade of transitions β → β'→ β'' are clearly demonstrated in the inset of Fig.1 using the VIN or BM EOS.

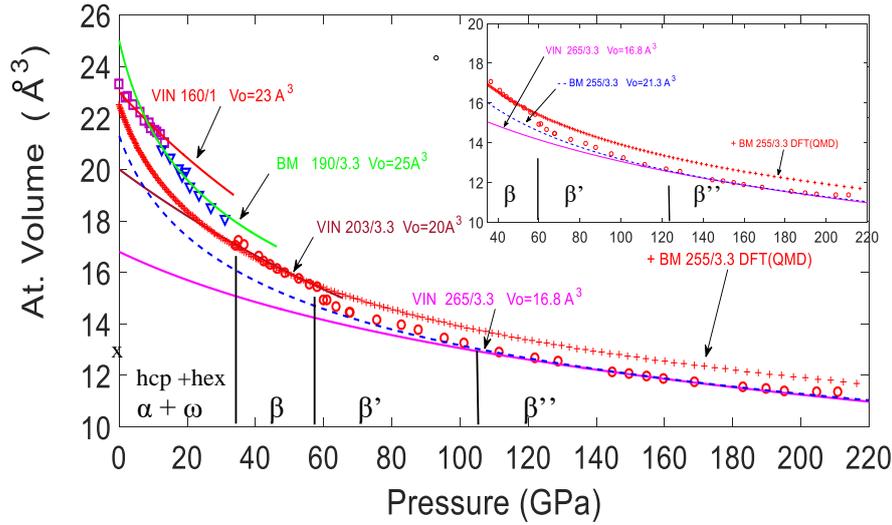

**Fig.1**: EOS of the elemental zirconium room temperature isotherm. The experimental data points are taken from Stavrou et al. [1] and are fitted with the VIN or BM EOS. The successive transformation from hcp to hexagonal to bcc structures are observed. The derived bulk moduli are assigned $B_o$,$B_o$' and $V_o$ pointed with arrows. Above 30GPa the β phase is dominant and a successive cascade to bcc-β' and bcc-β'' are clearly demonstrated. The bcc-β structure is stable up to ~58GPa, where a first order phase transition occurs with a volume reduction of ~5% which is stable up to ~105GPa (inset blue solid line). Above 105GPa and up to 220GPa a stable bcc-β'' exists (magenta solid line). The dashed blue line represents the BM simultaneous best fit of the experimental data in the P-V and P-T planes (see Fig.2) using the combined approach (see text). The red + sign is the DFT(QMD) simulation explained in the text.

Summery of the parametrized bulk moduli and the volumes at zero pressure are given in table I:

Table I. Summary of the elastic properties of Zr metal derived by VIN and BM equations of state. $V_o$ is obtained by extrapolation to zero pressure as shown in Fig.1 . Note the different scale of $V_o$ obtained with BM and VIN EOS.

| Zr Phase | $B_o$ (GPa) | $B_o'$ | $V_o$ (Å³/at.) | Fitting Procedure |
|---|---|---|---|---|
| α - hcp | 160(5) | 1.(0.3) | 23(2) | VIN |
| ω - hex | 190(5) | 3.3(2) | 25(2) | BM |
| β bcc | 203(7) | 3.3(2) | 20(2) | VIN |
| β' bcc | 230(7) | 2.0(2) | 18.3(3) | VIN |
| β" bcc | 265(5) | 3.3(1) | 16.8(4) | VIN simultaneously LG |
| β" bcc | 255(10) | 3.3(2) | 21.3(3) | BM dashed blue line Fig.1 |

The thermally corrected (actual pressure) experimental melting data reported by Prisiades et al. [7] is in accord with our calculations. However, the thermal contribution upon raising the temperature experienced by the Zr sample confined in the DAC, $Po_{th}$ and $Tm_o'$, are not taken into account. Thus, poor fittings of LG or Simon-Glatzel are obtained (there in Fig.5). Applying the present combined approach, namely the LG criterion using $B_o$ and $B_o'$ determined above, deriving $Po_{th}$ and $Tm_o'$ and adjusting $\gamma_o$ accordingly, reveal an excellent fit of the thermally corrected experimental melting points. This allows the extrapolation of the melting curve to high pressures and temperatures shown in Fig.2:

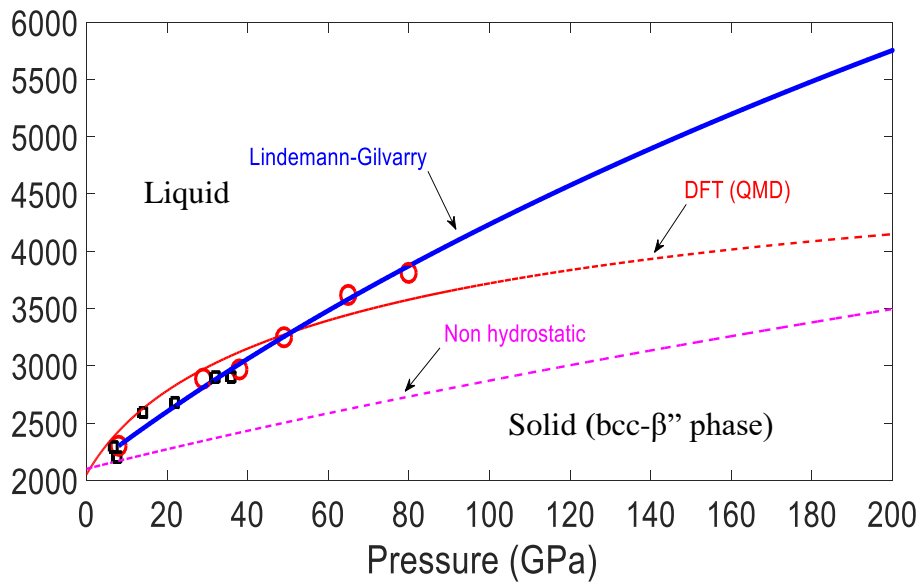

Fig 2. Melting curve of elemental Zr. The experimental melting data red points are the thermally corrected pressure values reported by Prisiades et al.[7]. The black squares are recent results by H.B. Padousky [7]. The solid blue line stands for BM and LG combined approach fitting procedure with the parameters $B_o$=255GPa and $B_o'$= 1.35 and $\gamma_o$=2.15. As isochoric condition exist in the DAC, the calculated thermal pressure shift $Po_{th}$= 8GPa and the melting temperature at $Po_{th}$, $Tm_o'$= 2300K, are the corrections needed for the Zr sample confined in the DAC. The red dashed line is the DFT(QMD) simulation combined with the LG criterion (see discussion). The magenta dashed line presents the nonhydrostatic thermodynamic formalism [28].

Choosing VIN or BM equation of states for fittings the experimental melting points within combined approach, both give nice fitting results, however, with different elastic parameters and different melting curve for elemental Zr.

Discussion –

The EOS elemental zirconium at ambient temperature (isotherm 300K) reported by Stavrou et al [1] was performed in a laser heated DAC were the PTM was fine powder of Zr, namely, no PTM was used. The influence of the PTM on the high pressure melting results was debated since the early studies using the DAC technique and is under discussion up to date [26]. The use of fine powder of Zr as PTM remove uniaxial stress and possible misleading EOS. These excellent data warrant a precise analysis which utilize both VIN or BM EOS as shown in Fig.1. The bulk moduli $B_o$, $B_o'$ are derived separately for each of the crystallographic phases α,ω, β, β' and β''. The β'' phase, fitted by both BM and VIN equations of state, showing that by no means β' phase can be denied. However, it can be argued that within the error bars, which are not depicted in Ref.1, the claimed β' phase do not exist. The analysis of any experimental point is referred to the center of mass of the error bar. What should be looked at is the trend of the experimental points relative to the fitting line. In the present case all points in the region from 60 to 100GPa are above the BM and VIN fitting lines of the β'' phase, namely both reveal the β' phase.

The fitting parameters, including the extrapolated $V_o$, are summarized in table I. Generally, best fit solutions are not unique and strongly depend on the chosen EOS and on the initial $B_o$ and $B_o$' parameters inserted in to the fitting program. Thus, the derived bulk moduli given in Table I indicate an alternative best solutions for the P-V equations of state. This specially applies to the elastic parameters proposed for α,ω,β, β' phases. As shown in Fig.1 the transition ω → β seems to be of first order type with volume collapse of ~10%. In the case of the β" phase the derived bulk moduli $B_o$,$B_o$' simultaneously best fit both, the EOS (Fig.1) and P-T phase diagram (Fig.2 ). Parisiades et al. [7] failed to fit their data with LG criterion because they used as an anchor the handbook ambient melting pressure (2128K) measured outside the DAC, namely not in confined environment. This melting point do not exist in the DAC. By introducing the thermally corrected $Po_{th}$ and $Tm_o$' and adjusting $γ_o$ a perfect fit is obtained, showing that the combined approach is reasonable. It should be mentioned that H. B. Radousky et al. recently published in PRB (2020) [29] utilizing the Lindemann criterion and ignoring the EOS, have totally missed the experimental data reported by Parisiades et al. [7]. The mismatch is clearly observed starting at 64-80GPa experimental points through the extrapolation to 160GPa yielding 1000K difference.

The same mismatch are the simulations performed by of L. Zhang et al. published in Nature (2019) [28] applying a new nonhydrostatic thermodynamic formalism for solid mediums.

The present analysis reveal a cascading bcc to bcc, pressure dependent volume transitions β → β' → β" in the pressure region 34-200GPa. The first order volume collapse β → β' is followed by a smooth transformation to the β" phase at ~135GPa, in contradiction to the claim made in [1]. The β" phase is the stable and is dominant above 110GPa at 300K for Zr metal and is the phase approaching the melt, briefly described by Parisiades et al. [7].

Stavrou et al. [1] applied first-principles combined density-functional theory (DFT) and finite temperature quantum molecular dynamics (QMD) calculations for the bcc Zr. Their results support the idea of a first-order pressure-induced isostructural phase transition which is triggered by un-harmonic motion. These simulations, as shown by the green line (there in Fig.1), do not pronounce the volume collapse cascades α → ω → β and the β → β'→ β". In addition, the present analysis reveal that the β' phase smoothly transform to β", hinting of a second order type. It could be argued that the QMD calculations do not represent the β" 300K isotherm but apparently a higher isotherm of β" which eliminates all the transitions like in the case of Ce [3].

The DFT(QMD) simulated melting curve depicted in Fig.2 was obtained by combining the P-V simulated data [1] with the LG approximation. The mismatch demonstrates the difficulties of the DFT theory to predict melting curves.

Utilizing the non-hydrostatic thermodynamic formalism for solid mediums, propose by Lin Zhang et al., published in Nature (2019), is shown in Fig.2 by the magenta dashed solid line. The absolute mismatch of this formalism calls for reconsideration of this formalism.

**Conclusions** – Analysis of the experimental isotherm 300K data of Zr metal EOS, separately for each identified crystallographic phase, reveal cascading crystallographic transitions β → β' → β". Upon increasing the pressure β' phase smoothly transforms to β" phase (at ~135GPa) indicating a moderate second order phase transition. The analysis of the phases α, ω, β, β' and β" show different bulk moduli $B_o$ and $B_o'$ and different zero pressure volumes ($V_o$) than those reported in the literature.

The present derived bulk moduli of the bcc-β"phase, are confirmed simultaneously by the approximated Lindemann-Gilvarry formula according to the procedure suggested in steps1 and 2 (combined approach). Inserting the actual pressure $Po_{th}$ and the actual melting temperature $Tm_o'$ and adjusting $\gamma_o$, yield a perfect fit of the reported melting data [7], yielding a reasonable extrapolation to high pressures and temperatures. The existence of the bcc-β' phase is here reported for the first time.

**Acknowledgements**

The superb high quality DAC experiment of E. Stavrou et al. [1] is highly appreciated.
The author gratefully acknowledge Prof. Z. Zinamon, Department of Particle Physics, Weizmann Institute of Science, Rehovot – Israel, for the many helpful illuminating discussions and comments.
Thanks are due to Dr. Lonia Friedlander, Ilse Katz Institute for Nanoscale Science and Technology, Ben-Gurion University of the Negev, Beer Sheva for many helpful comments and clarifying discussions.